\documentclass[aip,twocolumn,cha,showpacs]{revtex4}
\usepackage{amssymb}
\usepackage{epsfig}
\usepackage{graphicx}
\usepackage{amssymb}
\usepackage{wasysym} 

\begin{document}

\title{The Heads and Tails of Buoyant Autocatalytic Balls}

\author{Michael C. Rogers$^{1}$ and Stephen W. Morris$^{2}$}

\affiliation{$^{1}$Department of Physics, McGill University,  3600 rue University, Montr{\'e}al, Qu{\'e}bec, Canada H3A 2T8}

\affiliation{$^{2}$Department of Physics, University of
Toronto, 60 St. George St., Toronto, Ontario, Canada M5S 1A7}

\date{\today}

\begin{abstract}
Buoyancy produced by autocatalytic reaction fronts can produce fluid flows that advect the front position, giving rise to interesting feedback between chemical and hydrodynamic effects. In a large diameter, extended cylinder that is relatively free of boundary constraints,  localized initiation of an iodate-arsenous acid (IAA) reaction front on the bottom boundary generates a rising autocatalytic plume. Such plumes have several differences from their non-reactive counterparts. Using numerical simulation, we have found that if reaction is initiated using a spherical ball of product solution well above the bottom boundary, the subsequent flow can evolve much like an autocatalytic plume: the ball develops a reacting head and tail that is akin to the head and conduit of an autocatalytic plume, except that the tail is disconnected from the boundary.  In the limit of large initial autocatalytic balls, however, growth of a reacting tail is suppressed and the resemblance to plumes disappears. Conversely, very small balls of product solution fail to initiate sustained fronts and eventually disappear.
\end{abstract}


\maketitle

%
%
{\bf When feedback between autocatalytic chemical reaction and fluid flow occurs, the resulting ``chemo-hydrodynamics'' can lead to novel flow phenomena and interesting instabilities. The buoyancy produced by an autocatalytic reaction can drive fluid flows that deform the reaction front in an otherwise quiescent fluid. An extreme example of this deformation is the formation of rising plumes with complex morphologies and dynamics.  In this paper, we briefly review work on fluid flows driven by the iodate-arsenous acid (IAA) reaction, an archetypal cubic autocatalytic reaction.  We consider various geometries with an emphasis on the rich dynamics of autocatalytic plumes. We then present numerical simulations of the effect of buoyancy driven flow on initially spherical ``autocatalytic balls''. Evolution of the flow structure resulting from an autocatalytic ball exhibits different morphological regimes that depend on the initial size of the ball. Extremely small initial balls die away and fail to produce sustained reaction fronts or fluid flow. Over a certain range of initial radius, the autocatalytic ball grows to resemble the structure of an autocatalytic plume, with a well-defined head followed by an elongating tail. Beyond this regime, for a larger initial ball radius, a prolonged tail does not form, and the majority of reaction product remains confined within an ascending vortex ring.}

\section{Introduction}

The propagating front in the IAA reaction consumes solution that is in a state where essentially no reaction has occurred, and leaves in its wake a nearly fully reacted solution that is less dense than the reactant. Heat is produced during the reaction, as it is slightly exothermic. In addition, there is also an isothermal density change due to the difference in partial molal volumes of the reactant and product solutions. In the presence of gravity, a front configuration where less dense product solution is beneath denser reactant solution is hydrodynamically unstable to buoyancy driven convection.  Convection typically deforms the front and causes it to propagate more quickly than it would in the absence of flow.

\begin{figure}
\begin{center}
\includegraphics[width=8.2cm]{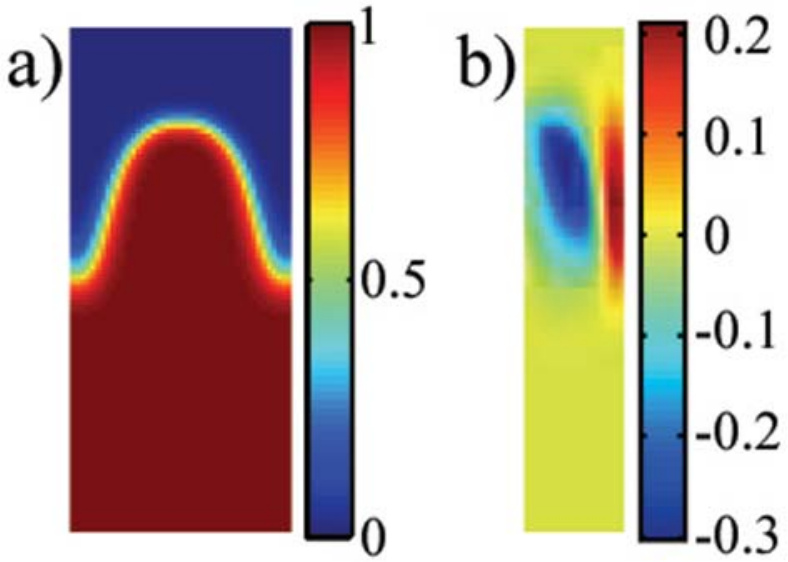}
\caption{
(color online) a) The axisymmetric dimensionless concentration field is reflected across the axis to show the shape of the reaction front for a tube with radius 20$\ell$. b) The dimensionless vorticity field for the same front. Negative vorticity corresponds to clockwise rotation.  A buoyancy-driven axisymmetric vortex ring is found in the vicinity of the front, with the fluid rising on the axis. The simulation used to generate this image is discussed in Section IV. $\ell$ is a length scale which is on the order of the front thickness.}
\label{cap_front}
\end{center}
\end{figure}

The buoyancy-enhanced propagation of the IAA front has been carefully studied in various geometries. In thin vertical tubes, the relative strengths of the buoyancy and viscous forces due to the confining walls of the tube control the stability of an ascending front.  The front shape  can be flat, asymmetric, or axisymmetric with respect to the tube axis~\cite{pojman1,vasquez_01,MASjpc}. When buoyancy is suppressed by a small tube radius, the front remains flat. At tube radii just above the onset of convection, the front becomes asymmetric. Further increasing the tube radius causes the front to develop an axisymmetric deformed shape, as shown in Fig~\ref{cap_front}, where the corresponding fluid flow is a vortex ring spanning the tube that moves with the front. We used this axisymmetric state as a test case for our numerical scheme, discussed in Section IV below.

Regardless of their morphology, all ascending fronts in tubes have the property that their shape remains constant as they propagate. Since the shape is constant, front propagation is essentially a one-dimensional motion, where 
the velocity of the front only depends on the relative buoyancy, the tube radius, and the reactant concentrations~\cite{hanna,MASjpc}. When a front is propagated against an imposed Poiseulle flow, its speed and shape also depend on the direction and magnitude of the imposed flow~\cite{Vasquez.PRE.07}. Surprisingly, in the wide gap limit for Poiseulle flow, there is almost no effect on the traveling velocity when reacting fronts travel in the opposite direction of the imposed flow~\cite{edwards,leconte}. 

In the quasi two-dimensional limit of thin vertical slots or horizontal layers, a convecting reaction front takes on a two-dimensional pattern. As in the case of thin tubes, front motion and morphology in a thin horizontal layer depends on the chemical concentrations of the reactants~\cite{hanna, GRIBjpc, HARjpc}. Simulations show that for a completely confined horizontal layer of solution, the propagation speed and deformation of the front depends on layer thickness~\cite{rongy_covered}. In simulations of solutions with a free surface, surface tension gradients across the front can induce capillary (Marangoni) flows~\cite{rongy_uncovered}.

Rising reaction fronts that span the width of a thin vertical slot (a Hele-Shaw cell) are unstable and will develop a wave-like pattern of fingers~\cite{huang,carey,bockmann}. The fingers emerge from an initially flat front through competition between buoyancy-driven instability, which acts to extend the fingers vertically, and diffusion, which smoothes the pattern~\cite{SHARPphy, KEENphy, MIKpre, MARTpre}. Scaling of the fingers depends on the dimensions of the slot and reaction parameters~\cite{dewit}. Amplification of the finger pattern can be achieved by matching the spacing of the the fastest growing mode with periodic variations in slot thickness~\cite{HORVprl}. Interestingly, descending fronts in a vertical slot can also have buoyancy-driven instabilities~\cite{DHEprl,DHEjfm}, even though the descending buoyant front is initially stably stratified.  The pattern results from the difference between molecular and thermal diffusivities.

In the quasi one- and two-dimensional scenarios, motion of the reaction front is severely limited by the boundaries of the reaction geometry. We now focus on reaction-driven flows that develop in a three-dimensional geometry where boundary constraint is minimal. We first discuss the behavior of autocatalytic plumes, and then compare them to simulations of a closely related system: autocatalytic balls.

\section{AUTOCATALYTIC PLUMES}
\label{plumes}

\begin{figure}
\begin{center}
\includegraphics[width=5cm]{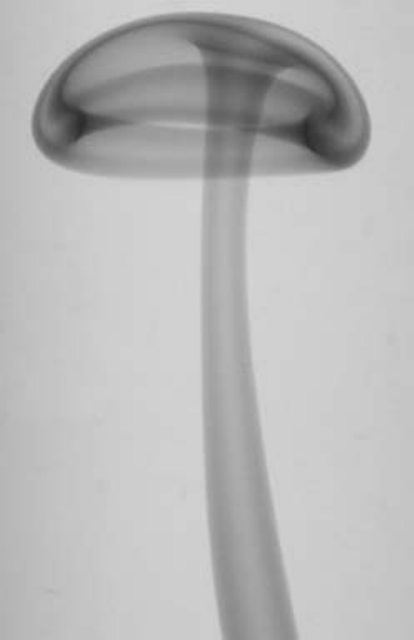}
\caption{
An autocatalytic plume produced by the IAA reaction. Starch indicator was used in solution to reveal the plume of product solution.}
\label{bw_plume}
\end{center}
\end{figure}

From the flow of air above a burning match to the rising molten rock that forms volcanic island chains, plumes at vastly different scales are abundant in Nature. Under laminar flow conditions, plumes consist of a well-defined plume head and a trailing conduit, as shown in Fig.~\ref{bw_plume}. Plume formation typically results from continuous forcing provided by a localized source of buoyancy. Usually, buoyancy is the result of a density difference caused by the thermal expansion of the fluid. Similarly, a source of fluid with a different chemical concentration will produce a plume due to the isothermal density difference of the fluid. Autocatalytic chemical reactions, such as the IAA reaction, produce both thermal and concentration density differences, and therefore can induce complex plume formation. 

The upwelling flow from a laminar plume conduit provides a continuous source of buoyancy to its head. It is usually the case that a plume head will remain attached to its conduit~\cite{GRIpfa, ROGpof}, which sustains its ascent and feeds a vortex ring flow within it. If a vortex ring detaches from its source and becomes ``free'', it is said to {\it pinch-off}. Free vortex rings are usually studied using a piston-cylinder arrangement~\cite{SHAarfm,GHAjfm}, where a single piston stroke is used to create the vortical flow. However, free vortex rings can also be created by the pinch-off of autocatalytic plume heads from their conduit~\cite{ROGprl,ROGpre2}. This process represents the most dynamically interesting stage of the evolution of an autocatalytic plume. These stages were described in detail in Refs.~\cite{ROGprl,ROGpre2}, and are summarized below.

Autocatalytic plumes can be formed by initiating the IAA reaction at the bottom of a vertically oriented capillary tube and letting the reaction front escape into the bottom of a much larger cylindrical tank containing a quiescent volume of reactant solution. Plume formation is initiated once the reaction front emerges from the submerged capillary tube. The earliest stage of plume evolution is by far the slowest, as the creeping flow of the tiny plume emerging from the capillary tube develops a small, roughly spherical head. The next stage involves the development of vortical motion in the head, where the upwelling product solution wraps itself toward the centre of the overturning vortex ring. At this point, the fate of the plume head --- whether or not it pinches off, and if so, how many times pinch-off is repeated --- depends on the fluid properties of the reactant solution. In cases where pinch-off occurs, the product solution flowing upwards in the conduit left behind forms a second generation plume head at the site where the conduit pinched. 

The evolution of autocatalytic plumes has been studied in fluids with various properties by altering the concentration of glycerol in the reactant solution~\cite{ROGpre2}. In general, pinch-off does not occur for solutions with low concentrations of glycerol. These solutions have the least viscosity and the fastest diffusion constants. At the other limit, pinch-off was seldom observed for the highest glycerol concentrations. However, pinch-off does occur reliably in an intermediate viscosity range for glycerol concentrations of 30\% and 40\% by volume. The highest number of pinch-offs observed for a given run was 5, for a 40\% glycerol solution. The number of pinch-offs observed during a given experimental run increased with the height of reactant fluid for a cylindrical tank with the same diameter, reflecting the geometric limitations of even very large experimental volumes. 

Although the external appearances of autocatalytic and non-reacting laminar plumes are similar, there are many important differences between them. A comparison of the steady-states reached by plume conduits was presented in Ref.~\cite{ROGpre}.  Autocatalytic plume conduits have reaction occurring along the entire interface between reacted and unreacted fluids, which causes the buoyancy flux to be distributed along the entire conduit length. This results in a velocity profile where the maximum velocity is near the interface between reacted and unreacted fluids. This contrasts the conduit velocity profile for a non-reactive plume, where the a maximum velocity is located along its axis of symmetry. The other major difference between the two types of conduits are their steady-state morphologies. Due to the ongoing  reaction along the front that forms the conduit, autocatalytic plumes have conical conduits that widen with distance from the base of the plume, whereas non-reacting laminar plume conduits are cylindrical. 

During the transient phase of plume growth, a non-reacting plume head is supplied with buoyancy by the upward flow in the attached conduit.  This flow must rise at a greater velocity than the upward speed of the head.  The plume head of a continuously driven, non-reacting plume is therefore unable to detach from its conduit. Moreover, after a short initialization period, the upward velocity of a non-reacting plume head is constant, and its dimensions maintain the same scaling relationship as it grows~\cite{ROGpof}. On the other hand, autocatalytic plume heads accelerate, pinch-off, and do not maintain self-similarity~\cite{ROGpre2}. The events leading to autocatalytic plume pinch-off were investigated in detail using simulation, and it was shown that a minimum velocity in the conduit beneath the head signals the onset of the pinching off process long before morphological indicators, such as narrowing of the conduit~\cite{ROGpre2}. Despite development of this velocity minimum, however, the head continues to accelerate upwards, demonstrating the lack of influence that the conduit has on the head at this stage of autocatalytic plume development. The ``conduit'' therefore no longer acts as a conduit: it does not transport buoyant fluid to the head. Instead, the fluid trailing the head is more like a ``tail'' being dragged along in its wake. In Section V.B below, we numerically explore the growth of such tails on ``autocatalytic balls" ---  buoyant, ascending spherical regions of fluid bounded by autocatalytic reaction fronts.   
  
\section{AUTOCATALYTIC BALLS}
\label{balls}

The simplest geometry for a three-dimensional reaction front is a sphere. In the absence of buoyancy-driven flow, spherical autocatalytic reaction fronts have previously been studied both theoretically and numerically. These studies have focused on the existence and stability of spherically symmetric solutions of the reaction-diffusion equations~\cite{JAKpre, TOTHpre}, and on the effect of autocatalyst decay~\cite{JAKpre2}. Threshold conditions needed to form spherical autocatalytic reaction fronts have also been calculated for both quadratic and cubic autocatalytic reaction fronts~\cite{needham}. 

Autocatalytic reaction fronts under gravity have some similarities to flame fronts~\cite{SIVcst, HORjcp}. Both systems involve positive feedback mechanisms that sustain chemical reactions occurring at their respective fronts. In the presence of an oxidizing agent, heat produced by exothermic chemical reactions in a combustible system sustains the reaction process. This process is analogous to the production of catalyst by an autocatalytic reaction front, which produces the catalyst needed to maintain propagation of its reaction front. As a consequence of this analogy, spherical autocatalytic reaction fronts are sometimes referred to as ``isothermal flame balls"~\cite{JAKpre, JAKpre2}. They have also been called ``spherical reaction balls"~\cite{TOTHpre}. In this paper we adopt the name {\it autocatalytic balls}.

Flame balls are steady, radially symmetric solutions of the reaction-diffusion-conduction equations for pre-mixed laminar flames. Originally proposed by Zel’dovich in 1944~\cite{zeldovich}, flame balls have since been observed when a spherical flame is formed under free fall conditions~\cite{RONcomb, RONaiaa}. The combustive process which sustains flame balls is highly exothermic, and therefore free fall conditions are necessary to prevent vigorous buoyancy driven flow. During flame ball experiments in microgravity, small local accelerations create unwanted gravity-like random accelerations called ``g-jitter". In the presence of these small acceleration fluctuations, flame balls are sometimes observed to deform into ``flame strings"~\cite{RONcomb, RONaiaa, BUCKcst}.  Flame strings are simply flame balls that have been stretched into a long, cylindrical shape. 

In addition to being caused by unwanted accelerations in microgravity experiments, the deformation of flame balls under gravity is particularly relevant to explosion scenarios for Type Ia supernovae~\cite{VLActm}. The initial stages of the march toward detonation of a Type Ia supernova involve the formation of a ``flame bubble" near the centre of a compact white dwarf star.  This reacting bubble is driven by its buoyancy toward the nearest surface of the star in the early stages of the supernova explosion, breaking its spherical symmetry. 

\section{AUTOCATALYTIC BALL SIMULATION}
\label{simulation}

To simulate the behavior of autocatalytic balls, we employed a dimensionless model previously used to study autocatalytic plumes~\cite{ROGpre2}. The model uses the semi-implicit method for pressure linked equations (SIMPLE) algorithm~\cite{PATbook, FERZbook} to calculate the relevant physical fields. The formulation of the model is detailed in this section.

The dimensionless parameters required for simulating autocatalytic balls come directly from the scaling of the basic dynamical equations. The thermal, chemical, and mechanical properties of the fluid are characterized by the Schmidt number
\begin{equation}
{\rm Sc}=\nu/D~, 
\label{schmidt}
\end{equation}
and the Lewis number
\begin{equation}
{\rm Le} = {\kappa}/{D}~, 
\end{equation}
where $\nu$ is the kinematic viscosity, $\kappa$ is the thermal diffusivity, and $D$ is the molecular diffusivity of the autocatalyst, which for the IAA reaction is the iodide ion. The ratio of ${\rm Sc}$ and ${\rm Le}$ is the Prandtl number, 
\begin{equation}
{\rm Pr} = \nu/\kappa~, 
\end{equation}
which is a useful quantity for characterizing purely thermal plumes~\cite{TURNER}.

For a 40\% glycerol IAA solution, using $\kappa = 1.2 \times 10^{-3}$~cm$^2$/s and values from Ref.~\cite{ROGpre2}, we find
\begin{equation}
{\rm Sc} = 9000 { , } ~~{\rm Le} = 280 {,} ~~{\rm and} ~~{\rm Pr} = 32. 
\label{lewis}
\end{equation}
These values of ${\rm Sc}$ and ${\rm Le}$ indicate that the rates of diffusion of heat and momentum are much faster than that of  concentration, implying  that the concentration profile at the reaction front has by far the steepest gradient of the relevant fields. 

The Boussinesq approximation was used to quantify the density change from the reaction as   
\begin{equation}
\rho=\rho_0[1-\alpha \Delta T - \beta \Delta c],
\label{boussinesq}
\end{equation}
where $\Delta T = T-T_0$ and $\Delta c = c-c_0$ are the temperature and concentration differences across the reaction front, and $\rho_0$, $T_0$, and $c_0$ are the initial conditions for the density, temperature and concentration, respectively. Here, $\alpha$ is the thermal expansion coefficient and $\beta$ is the compositional expansion coefficient. Since the density change comes from a linear combination of temperature and concentration effects, these contributions can be separated into two Rayleigh-like numbers~\cite{DHEprl, DHEjfm} 
\begin{eqnarray}
{\rm Ra}_T &=& \frac{g \alpha \ell^3}{\nu D} \Delta T ~~{\rm and}~~~{\rm Ra}_c = \frac{g \beta \ell^3}{\nu D} \Delta c
\end{eqnarray} 
where $\ell = \sqrt{D\tau}$ is the molecular diffusion length scale of the front, and $\tau$ is a time scale that depends on the front velocity $v_f$, such that $v_f=\sqrt{D/\tau}$.  The simulations in this paper used
\begin{equation}
{\rm Ra}_T=0.10 ~~{\rm and} ~~ {\rm Ra}_c=0.17~,
\end{equation}
which were calculated using the parameter values for an IAA solution containing 40\% glycerol given in Ref.~\cite{ROGpre2}.

Autocatalytic balls were assumed to be axisymmetric about the direction of gravity, allowing the simulation to be carried out in the cylindrical coordinates $(r,z)$, where gravity is oriented in the negative $z$ direction. In this coordinate system, the set of dimensionless equations used to model autocatalytic ball evolution were
\begin{equation}
\mathbf{\nabla} \cdot \mathbf{u} = 0,
\label{continuity}
\end{equation}
\begin{equation}
\frac{1}{\rm {Sc}} \frac{D\mathbf{u}} {Dt} = - \mathbf{\nabla} p + \mathbf{\nabla} ^2 \mathbf{u} + (Ra_{T}T+Ra_{c}c)\mathbf{e}_{z},
\label{NSE}
\end{equation}
\begin{equation}
\frac{Dc} {Dt} = \mathbf{\nabla}^2c + F(c),
\label{cEQN}
\end{equation}
and
\begin{equation}
\frac{DT} {Dt} = {\rm Le}\mathbf{\nabla}^2T + F(c).
\label{TEQN}
\end{equation}
Here, $\mathbf{e}_{z}$ is a unit vector pointing in the $z$ direction and $D/Dt$ is the material derivative. 
The cubic autocatalytic reaction term is
\begin{equation}
F(c)= c^2(1- c)~,
\end{equation}
where the dimensionless concentration $c$ is normalized by the maximum possible concentration of the product solution. This normalization scheme gives a maximum dimensionless concentration of $c=1$ and a minimum of $c=0$. The dimensionless temperature $T$ is  normalized in a similar way as the concentration, and is given by scaling the physical $\Delta T$ by $\Delta T_H = - \Delta H \Delta c/\rho_0 C_p$, which bounds it by $0 \le T \le 1$. Here, $\Delta H$ is the heat of reaction, and $C_p$ is the specific heat at constant pressure. 

The cylindrical boundaries of the system are given by $r=r_b$,  $z=0$ and $z=z_b$. Conditions at the boundaries are defined by imposing no flow of concentration or temperature across the boundary, and no-slip, giving
\begin{eqnarray}
{\bf u}&=&\frac{\partial T}{\partial r}=\frac{\partial c}{\partial r}=0~~~{\rm at}~~~r=r_b,\\
{\bf u}&=&\frac{\partial T}{\partial z}=\frac{\partial c}{\partial z}=0~~~{\rm at}~~~z=0~~{\rm and}~~ z=z_b.
\end{eqnarray}
The rather small length $\ell$, which is of order the thickness of the reaction front, is used to scale all of the other lengths in the model.  This means that rather large dimensionless values of $r_b$ and $z_b$ are required to model an extended volume of fluid.

\begin{figure}
\begin{center}
\includegraphics[width=8.9cm]{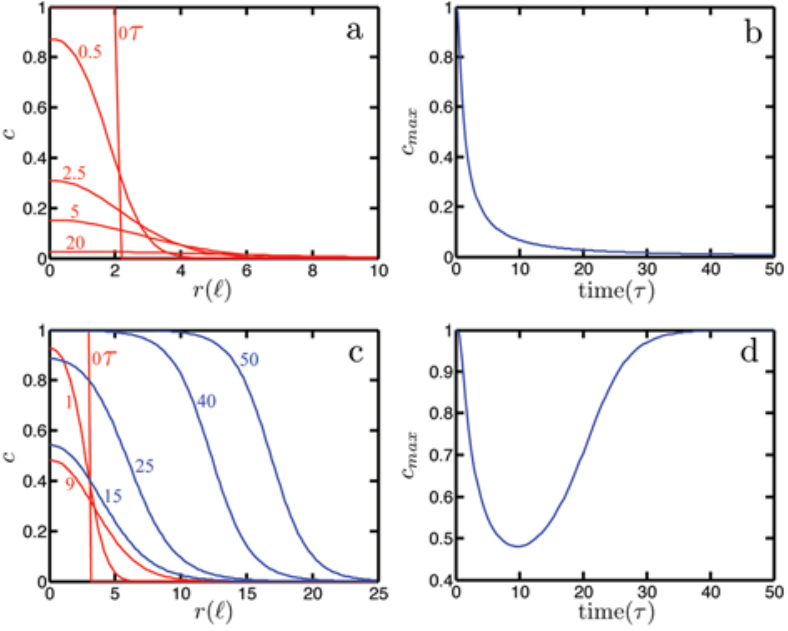}
\caption{
(color online) a) For an autocatalytic ball with $r_0=2\ell$, the concentration profiles at different times, in units of $\tau$, and b) the change in the maximum value of the concentration $c_{max}$  with time. c) The concentration profiles at different times for an autocatalytic ball with $r_0=3\ell$. Red curves indicate concentration profiles when $c_{max}$ is decreasing, blue curves are profiles when $c_{max}$ is increasing. d) The change in $c_{max}$ over time for the $r_0=3\ell$ ball.}
\label{death_all}
\end{center}
\end{figure}

\section{SIMULATION RESULTS}

\subsection{Front death}

Small localized regions of autocatalyst are not capable of sustaining three-dimensional reaction fronts for an autocatalytic reaction with cubic kinetics~\cite{needham}. This ``front death'' phenomenon is caused by diffusion, and is unlike the threshold behavior of excitable systems, such as the action potential of neurons~\cite{FITZbj} or the propagating oxidation wave in the Belousov-Zhabotinsky (BZ) reaction~\cite{ZAIKnat}. Neurons and the BZ reaction are excitable in the sense that they remain in a stable state until a threshold is reached, sending them into an unstable state. For a neuron, excitation occurs once a certain voltage, or action potential, is reached, resulting in the transmission of an electric signal. The excited state of the BZ reaction is reached when enough catalyst is present to transmit a propagating oxidation pulse through the solution. A BZ reaction pulse does not consume all of the reactants as it moves through the reactant medium, allowing for multiple excitations. A cubic autocatalytic reaction like the IAA reaction, in contrast, always has locally unstable kinetics~\cite{needham}. Therefore the presence of {\it any} amount of catalyst will render the solution unstable. However, despite the localized instability created by the presence of autocatalyst, three-dimensional diffusion can prevent the initiation of a reaction front provided that a critical amount of catalyst has not been exceeded~\cite{fb_foot}. In this section, the consequences of front death on autocatalytic balls is explored using our numerical model. 

To initiate autocatalytic balls, spheres located on the axis of symmetry of the coordinate system with $c=1$ and various initial radii $r_0$ were used as initial conditions. Convective effects were suppressed in the simulations exploring front death, so that it could be studied as a pure reaction-diffusion phenomenon. Even when buoyancy is introduced into the calculation, the buoyant force is negligible on small spheres in the radius range where the front dies, and very little convective transport occurs during the time scale it takes for the front to die. All simulations investigating front death were performed on a spatial domain of $z_b=r_b=120\ell$, with the centre of the initial ball located on the axis of symmetry at $z_0=60\ell$.

The radial distribution of product solution at different times for an autocatalytic ball with $r_0=2\ell$  is shown in Fig.~\ref{death_all}a. The time evolution shows that the product concentration simply diffuses away from its initial position without forming a reaction front. The maximum value of the concentration, $c_{max}$, is shown for the same ball in Fig.~\ref{death_all}b. The decrease in $c_{max}$ towards $c=0$ indicates death of the reaction front. An autocatalytic ball with an initial radius of $r_0=3\ell$ has a different fate. In this case, the initial front begins to diffuse away as it did for $r_0=2\ell$. After a period of time, however, a sustained reaction front is initiated and the autocatalytic  ball recovers from the initial concentration drop. This is seen in the radial distribution of product concentration for various times shown in Fig.~\ref{death_all}c, and the evolution of the maximum concentration shown in Fig.~\ref{death_all}d. At times less than $\sim10\tau$ the concentration diffuses and insufficient new product is generated. After that time, however, the front recovers, leading to the stable outward propagating front structure shown in Fig.~\ref{death_all}c at $40\tau$ and $50\tau$.

\subsection{Heads and tails}

When buoyancy is suppressed and front death does not occur, autocatalytic balls maintain a spherical shape as they expand. We now turn our attention to the evolution of autocatalytic balls under gravity. This involves an interesting dynamic interplay between reaction, diffusion, and buoyancy-driven flow. For these simulations, only autocatalytic balls that do not die are considered. 

\begin{figure}
\begin{center}
\includegraphics[width=8.2cm]{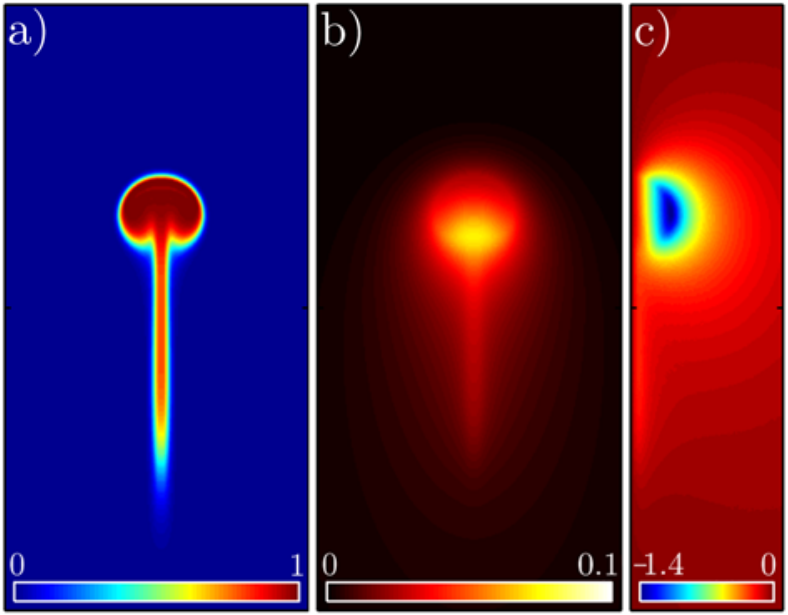}
\caption{The a) concentration, b) temperature, and c) vorticity fields at time $20\tau$ for an ascending autocatalytic ball with $r_0=15\ell$. The spatial domain of the simulation was $r_b=500\ell$ and $z_b=2000\ell$, however only data within the coordinates $r \leq 100\ell$ and $200  \ell \leq z \leq 600 \ell$ is shown. Concentration and temperature data are shown reflected across the $z$ axis, vorticity data, where negative values indicate clockwise motion, is not reflected.}
\label{all_fields}
\end{center}
\end{figure}

An example of the concentration, temperature, and vorticity fields from a simulation of an autocatalytic ball with $r_0=15\ell$ are shown in Fig.~\ref{all_fields}. The large ratio of the diffusivities $\kappa$ and $D$ ({\it i.e.} the large Lewis number Le) is evident in the difference between the concentration and temperature fields, the temperature field being the more diffuse of the two. The structure of the vorticity field in Fig.~\ref{all_fields}c is typical of the ring vortex that develops in all ascending autocatalytic balls. 

\begin{figure}
\begin{center}
\includegraphics[width=7.9cm]{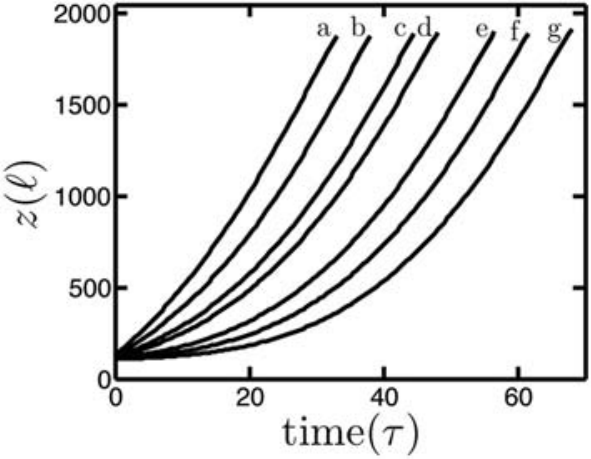}
\caption{
The ascent of autocatalytic balls with different initial radii over time. The curves correspond to autocatalytic balls with initial radii $r_0$  of a) 27.5$\ell$, b) 22.5$\ell$, c) 17.5$\ell$, d) 15$\ell$, e) 10$\ell$, f) 7.5$\ell$, and g) $5 \ell$.}
\label{fb_heights}
\end{center}
\end{figure}

The location of the top edge of the concentration jump for  autocatalytic balls with different initial radii $r_0$ is shown in Fig.~\ref{fb_heights}. The centres of the balls in these simulations were initially located at $z_0=100 \ell$ in a spatial domain with $r_b=500 \ell$ and $z_b=2000 \ell$.  The simulation time-step was $\Delta t=0.005 \tau$. These are the same parameters used in Ref.~\cite{ROGpre2} to simulate autocatalytic plumes, with the exception of the initial ball location $z_0$, which was farther from the lower boundary of the computational domain than the one used for plumes. The autocatalytic ball with the largest initial radius $r_0$ reaches the top of the computational domain faster than the others. The smaller the $r_0$ of the ball, the longer it takes to reach the top. All of the balls reach a terminal velocity that is approximately the same, increasing only slightly with $r_0$. The range of terminal velocities was 64.9 to 68.9 in units of $\ell/\tau$, which corresponds to a physical velocity difference of only $5 \times 10^{-3}$~cm/s between the smallest and largest $r_0$. The terminal velocities are dictated by the final amount of product solution in the ascending flow structure. As the small range of terminal velocities suggests, the amount of reacted fluid in the vortex ring once it reaches the top boundary is similar for all values of $r_0$. The integral of the concentration field for the vortex ring indicates that the amount of product reaching the top for the $r_0=5 \ell$ and $r_0=27.5 \ell$ balls was different by only $\sim 8\%$, with the larger ball delivering a slightly larger amount of product. Even though a larger $r_0$ balls starts with a larger amount of product, the slow initial ascent of the smaller ones gives them more time to expand and accelerate before reaching terminal velocity of their way to the top.

\begin{figure}
\begin{center}
\includegraphics[width=8.2cm]{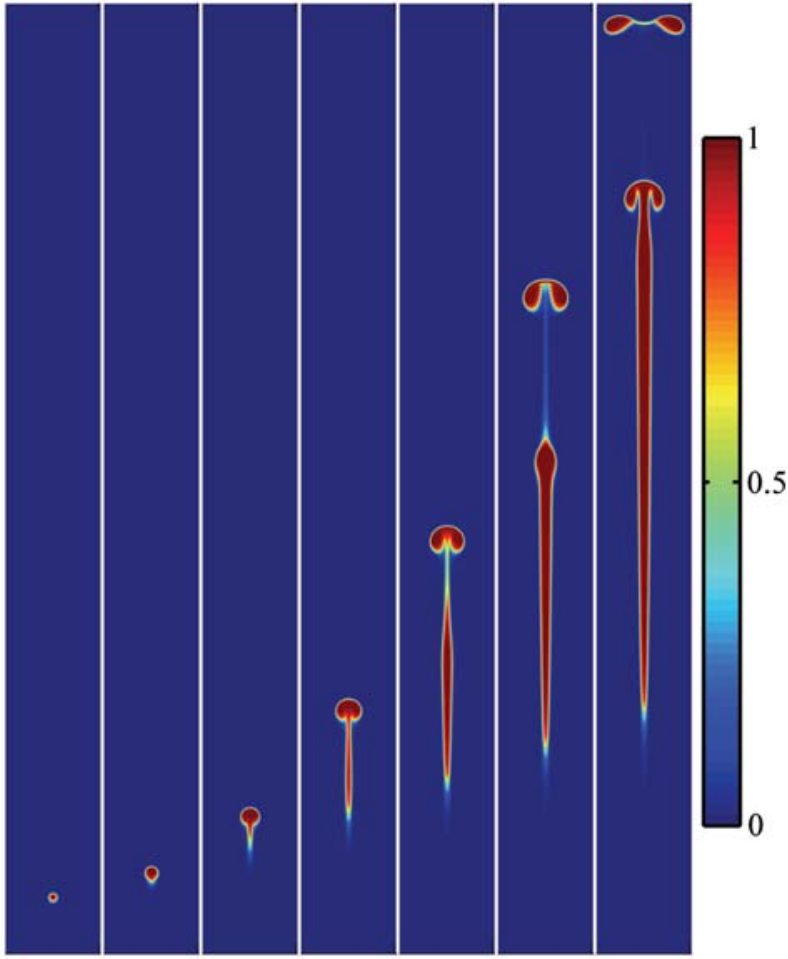}
\caption{
The evolution of an autocatalytic ball with $r_0=5\ell$, showing the development of the head and a tail of reacting solution. The spatial domain of the simulation is $r_b=500\ell$ and $z_b=2000\ell$, however the portion of the field displayed field shown has a radius of $100\ell$. From the left-most frame to the right-most frame, the elapsed time after initiation is $10\tau$ to $70\tau$ in increments of $10\tau$.}
\label{string}
\end{center}
\end{figure}

The evolution of the morphology of a small autocatalytic ball with $r_0=5\ell$ is shown in Fig.~\ref{string}. In the initial stages of its evolution, it remains roughly spherical. However, as it ascends it leaves behind a tail: a thin trail of product solution extending into the wake of the rising autocatalytic ball. As the tail elongates, the upper portion of the ball develops into a vortex ring that resembles a plume head. Just as an autocatalytic plume head pinches off from its conduit, the autocatalytic ball head pinches off from its tail. As it does so, the top of the tail grows a new, second-generation head, and the first-generation head rises upwards as an essentially free vortex ring. 

\begin{figure}
\begin{center}
\includegraphics[width=8.2cm]{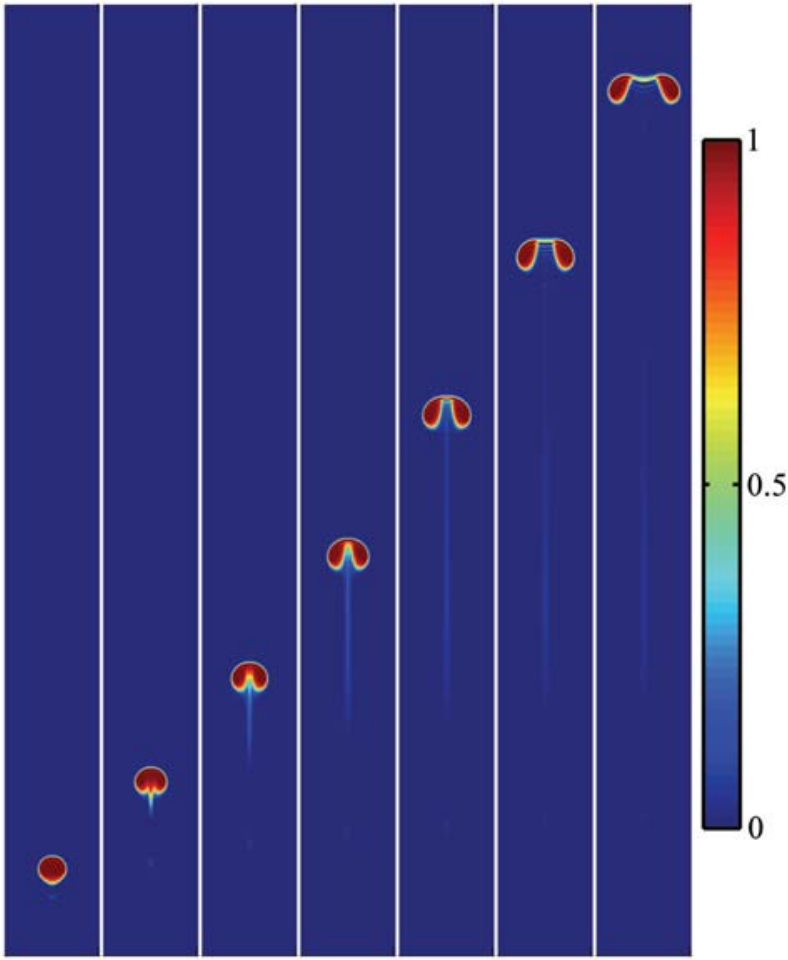}
\caption{
The evolution of an autocatalytic ball with $r_0=27.5 \ell$, showing that only a very thin filament of reacted solution trails the head, and that the filament is not capable of producing a sustained reaction front. The spatial domain of the simulation is $r_b=500 \ell$ and $z_b=2000 \ell$, however the portion of the field displayed has a radius of $100\ell$. From the left-most frame to the right-most frame, the elapsed time after initiation is $2.5\tau$ to $32.5\tau$ in increments of $5\tau$.}
\label{nostring}
\end{center}
\end{figure}

The similarity between the morphological evolution of autocatalytic plumes and balls disappears for balls with larger initial radii.   A ball with  $r_0=27.5 \ell$,  which fails to develop a reacting tail, is shown in Fig.~\ref{nostring}. As it rises, the ball leaves behind a filament of reacted solution. However, unlike the tails formed in the wake of autocatalytic balls with $r_0 \le 25 \ell$, the filament is so thin and dilute that it diffuses away before it can initiate a reacting front, much like how a very small autocatalytic ball undergoes front death. While the exact cutoff radius for reacting tail development, $r_{0t}$, was not determined, the simulation results show that it lies in the range $25 \ell < r_{0t} \le 27.5 \ell$. 

To help understand why tail production ceases once $r_0$ surpasses $r_{0t}$, it is useful to use a dimensional analysis argument for autocatalytic balls similar to one developed for astrophysical flame bubbles~\cite{VLActm}. For highly viscous autocatalytic balls, two time scales can be compared to estimate the conditions when a spherical ball of product undergoes significant deformity from viscous effects. These time scales are the viscosity time scale
\begin{equation}
\tau_v=\frac{r_0^2}{\nu},
\end{equation}
and the buoyancy time scale
\begin{equation}
\tau_b=\sqrt{\frac{r_0}{g'}},
\end{equation}
where $g'=g(\rho_u - \rho_r)/\rho_u$ is the buoyancy of the flame ball, with $g$ being the gravitational acceleration, and $\rho_r$ and $\rho_u$ being the densities of the reacted and unreacted fluid, respectively. The ratio of the two time scales
\begin{equation}
\frac{\tau_b}{\tau_v} = \frac{\nu}{\sqrt{g'}}r_0^{-3/2}
\end{equation}
shows which time scale dominates as a function of the initial radius $r_0$. For the case where $\tau_b\ll \tau_v$, the ball travels upwards for a significant period of time before viscous effects to are able to alter its shape. The buoyant and viscous timescales are equivalent at a ball radius of  
\begin{equation}
r_{c}=\left ( \frac{\nu^2}{g'} \right)^{1/3}~.
\end{equation}
Using the parameters of the simulation, and a total density jump across the front of $\Delta \rho = (\rho_u - \rho_r)=6.4\times 10^{-4}$~g/cm$^3$~\cite{foot_props}, we find $r_c \sim 33\ell$. This scale is in rough agreement with the observed crossover $r_0$ between autocatalytic balls with reacting tails and those without.

\begin{figure}
\begin{center}
\includegraphics[width=8.4cm]{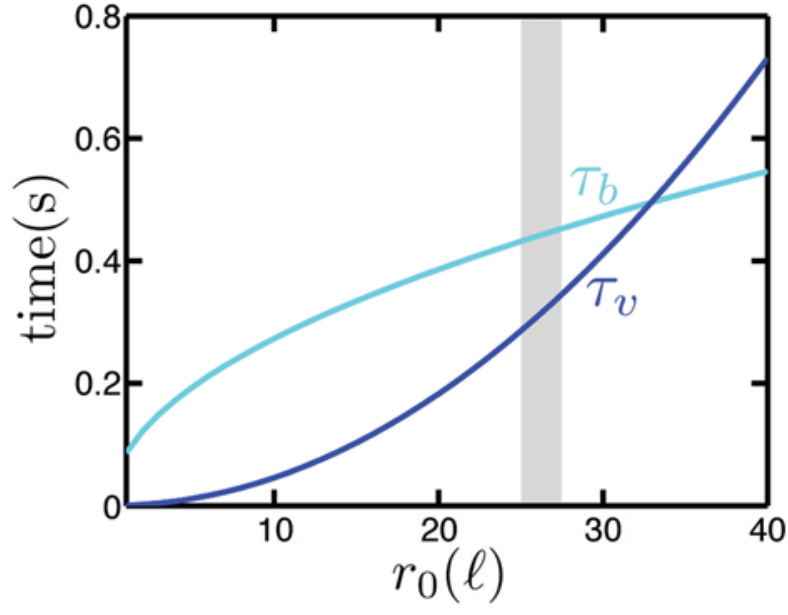}
\caption{
The viscous time scale, $\tau_v$, and the buoyancy time scale, $\tau_b$, as a function of autocatalytic ball radius $r_0$. The shaded area is the region where the transition from balls with tails to those without is observed in the simulations.
}
\label{fb_def}
\end{center}
\end{figure}

A plot of $\tau_b$ and $\tau_v$ in the range of $0 < r_0 \le 40 \ell$ is shown in Fig.~\ref{fb_def}. For $r_0 < r_c$, the timescales are such that $\tau_v < \tau_b$, and viscous forces have more time to act on, and therefore deform, the autocatalytic ball as it ascends. For  $r_0 > r_c$, $\tau_b$ becomes shorter than $\tau_v$ and viscous forces have less influence on autocatalytic ball morphology before the buoyancy forces carry the ball upwards. In our simulations, viscous deformation of the autocatalytic balls for $r_0\le 25\ell$ was capable of dragging enough product out of the autocatalytic ball to form a tail with a sustained reaction front of its own. For $r_0 \ge 27.5\ell$, however, buoyancy forces dominate and viscous effects drag only a small amount of reacted solution into the wake of the rising ball. The resulting filament of product solution diffuses away and is insufficient to initiate a reacting autocatalytic front.

\section{Conclusions}

Using a numerical scheme previously employed to simulate autocatalytic plumes, the evolution of buoyant autocatalytic balls was explored. Using reaction conditions for the iodate-arsenous acid reaction in 40\% glycerol solution, three different regimes of behavior for autocatalytic balls with different initial radii were found. Very small autocatalytic balls that undergo front death occupy the first regime.  Buoyancy is negligible in this case. In the second regime, where $3 \ell \leq r_0 \leq 25 \ell$, autocatalytic balls closely resemble autocatalytic plumes: they develop reacting heads and tails.  Pinch off can occur, producing secondary heads.  In the third regime, for the largest initial condition examined with $r_0 = 27.5 \ell$, only a small amount of product is dragged into the wake of the head, and the resulting filamentous structure does not produce a reacting tail.   A dimensional analysis argument involving the viscous and buoyancy time scales was used to help explain the disappearance of the tail. 

The scope of our understanding of autocatalytic plumes and balls has thus far been limited to those formed by the iodate-arsenous acid reaction. It would be interesting to explore how these phenomena behave in other autocatalytic reactions, for instance in a reaction in which the thermal and concentration contributions to buoyancy have opposite signs. The resulting interplay between chemical reaction and fluid flow would almost certainly produce interesting and surprising dynamical phenomena.  
~~

\begin{acknowledgments}
This work was made possible by the facilities of the Shared Hierarchical Academic Research Computing Network (SHARCNET:www.sharcnet.ca) and Compute/Calcul Canada. This research was supported by the Natural Science and Engineering Research Council (NSERC) of Canada.
\end{acknowledgments}


\begin{thebibliography}{10}
\bibitem{fb_foot}
{\rm Front death depends on the number of spatial dimensions and the reaction kinetics. Quadratic kinetics do not result in front death for any space dimension. For cubic kinetics, fronts always form in one and two space dimensions, but may exhibit front death in three dimensions.  See Ref.~\cite{needham} for a detailed theoretical treatment. }
\bibitem{foot_props}
{\rm The total density difference between a reacted fluid with $c=T=1$ and $c=T=0$ is $\Delta \rho=\Delta \rho_c+\Delta \rho_T$, where $\Delta \rho_c=4 \times 10^{-4}$~g/cm$^3$ is the isothermal density difference from compositional change, and $\Delta \rho_T=2.4 \times 10^{-4}$~g/cm$^3$ is the density difference caused by heating the reactant solution by 0.5 K.}
\end{thebibliography}
\end{document}